\documentclass[twocolumn]{aastex631}

\usepackage{amsmath}
\usepackage{amssymb}

\journalinfo{}

\makeatletter
\@ifundefined{frontmatter@title@above}{}{\def\frontmatter@title@above{\addvspace{1pt}}}
\@ifundefined{frontmatter@title@below}{}{\def\frontmatter@title@below{\addvspace{7pt}}}
\@ifundefined{frontmatter@preabstractspace}{}{\def\frontmatter@preabstractspace{0.15\baselineskip}}
\@ifundefined{frontmatter@postabstractspace}{}{\def\frontmatter@postabstractspace{0.15\baselineskip}}
\@ifundefined{frontmatter@above@affilgroup}{}{\def\frontmatter@above@affilgroup{\addvspace{1pt}}}
\@ifundefined{frontmatter@above@affiliation}{}{\def\frontmatter@above@affiliation{}}
\@ifundefined{frontmatter@above@affiliation@script}{}{\def\frontmatter@above@affiliation@script{}}
\makeatother

\newcommand{\msun}{M_{\odot}}
\newcommand{\mbh}{M_{\rm BH}}
\newcommand{\msp}{M_{\rm spike}}

\newcommand{\kms}{\,\mathrm{km\,s^{-1}}}

\shorttitle{Population search for dark matter spikes in megamaser rotation curves}
\shortauthors{Herrera}

\newcommand{\dM}{\Delta M_{\rm sp}}
\newcommand{\dlnl}{\Delta(-2\ln\mathcal{L})}

\begin{document}

\title{\large Population search for dark matter spikes in megamaser rotation curves}

\author{Gonzalo Herrera}
\affiliation{Kavli Institute for Astrophysics and Department of Physics, Massachusetts Institute of Technology,
77 Massachusetts Avenue, Cambridge, MA 02139, USA}
\affiliation{Harvard Laboratory for Particle Physics and Cosmology and Department of Physics, Harvard University,
17 Oxford Street, Cambridge, MA 02138, USA}
\email{gonzaloh@mit.edu}

\begin{abstract}
A black hole growing adiabatically inside a dark matter halo is expected to develop a dense spike, with a power-law slope between 1.5 and 2.5, depending on redshift and baryonic activity. Direct dynamical tests of this prediction are almost non-existent. Here we show that water megamaser disks offer one; their masers orbit within a parsec of the black hole and are mapped individually with very long baseline interferometry, so the enclosed mass at each radius can be read off the rotation curve. We develop a search method and apply it to eleven megamaser disks with public kinematics. Compared with a black hole alone, a spike is preferred in NGC~1194 and NGC~4258 and moderately improves the fit in several others. Allowing the disks to warp removes this preference in all but NGC~4258 and NGC~6264. No combination of disks, warped or not, allows a spike heavier than about ten per cent of the black hole mass, a sensitivity validated with mock injection tests. Mapping the warps with maser accelerations and detecting fainter inner masers, where a spike and a tilt differ most, can break the remaining degeneracy. Megamaser disks open a window onto dark matter around supermassive black holes, and, importantly, one that enables population statistics.
\end{abstract}

\section*{}

\begin{figure}[!b]\centering\includegraphics[width=0.87\columnwidth]{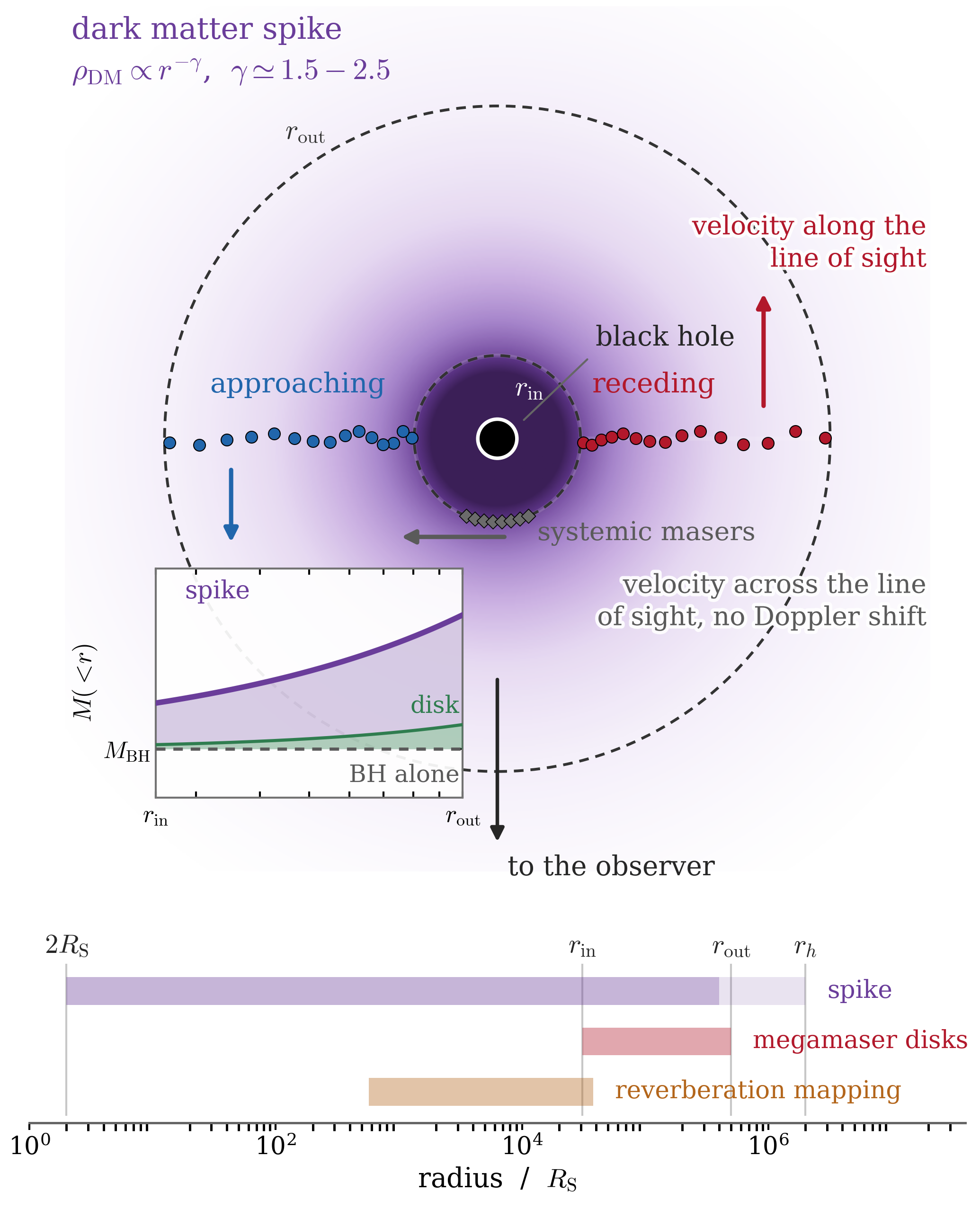}\caption{Illustration of a maser disk from above its plane, with a dark matter spike as a density field, and the masers on the midline. The inset shows the enclosed mass as a function of radii, and the lower bar shows the radial reach of this work, and reverberation mapping in \citet{sharma25}.\label{fig:schematic}}\end{figure}

Dark matter makes up about a quarter of the energy content of the Universe and shapes the growth of structure on almost every scale we have observed \citep{rubin70,clowe06,planck20,bertonehooper18}. What it does in the immediate surroundings of a supermassive black hole is much less clear. If a black hole grows slowly at the centre of a dark matter cusp, the deepening gravitational well pulls the dark matter inward and concentrates it into a steep ``spike'' \citep{gondolosilk99,quinlan95}, whose density falls with radius as $r^{-\gamma}$ with $\gamma$ between 2.25 and 2.5. Later encounters with stars soften the spike toward $\gamma\sim1.5$ \citep{gnedinprimack04,merritt04,bertonemerritt05,merritt07,vasiliev07,shapiroshelton16,herrera26,sharpe26,karydas26}, and a black hole formed from a massive seed in a baryonic environment, that was not born at the very centre, or that experienced a recent merger with nearly equal mass, would leave a shallower or truncated profile \citep{ullio01,merritt02,caiozzo25,herrera26}. A relativistic treatment of the growth moves the inner edge of the spike toward the black hole and raises the density there, more so around a spinning hole \citep{sadeghian13,ferrer17}. 

Finding a dark matter spike would be groundbreaking. It would be the first direct dynamical detection of dark matter on sub-parsec scales, a thousand times closer to the centre of a galaxy than rotation curves and lensing reach and where the density of cold dark matter should be highest, and its slope would record how the black hole grew and how stars and mergers have since reshaped its surroundings. It would also bear on the particle nature of dark matter: a spike of known mass fixes the expected annihilation signal, which scales as the square of the density \citep{gondolosilk99,bertonemerritt05}; self-interactions would reshape the spike and its inner halo \citep{shapiropaschalidis14,alonsoalvarez24,mezghanni26}; and the scattering of cosmic rays, neutrinos and photons off the dark matter around active black holes tests interactions inaccessible to laboratory experiments \citep{wang22,cline23b,ferrer23,herreramurase24,demarchi25}.

Direct tests are scarce. In the Galactic Centre, the orbit of the star S2 around Sgr~A* shows no precession beyond that of general relativity, which limits any extended mass inside the orbit to a small fraction of a per cent of the black hole and excludes steep spikes there \citep{lacroix18,shen24,gravity24}. The supermassive black-hole binary candidate OJ~287 has been argued both ways: the decay of its orbit has been read as evidence for a spike with $\gamma\approx2.35$ \citep{chan24} and as consistent with general relativity alone, which bounds $\gamma$ to about 2 or less \citep{alachkar22,deb25}; and X-ray binaries have given hints of spikes around stellar-mass black holes \citep{chanlee23}. 

None of these gives a population. Reverberation mapping of active galactic nuclei opened that route: tracing the enclosed mass with broad emission lines at several radii in fourteen active nuclei, \citet{sharma25} carried out the first population-level search for spikes and found hints of extended mass consistent with spikes of slope $\gamma\approx1.6$ in five of them. Reverberation masses, however, rest on a virial coefficient that is calibrated only on average, and that coefficient enters every inferred spike parameter as a multiplicative unknown; an independent, geometric measurement of the enclosed mass around a population of black holes is the natural next step.

Water megamaser disks avoid that coefficient. In some Seyfert nuclei, 22~GHz masers trace a thin, nearly edge-on disk of gas orbiting the black hole at radii of 0.02 to 1 parsec, and very long baseline interferometry maps each maser to tens of microarcseconds \citep{herrnstein96,kuo11,reid09}. The masers on the near and far sides of the disk move straight toward and away from us, so their line-of-sight velocities are their orbital velocities, and the mass enclosed inside each maser's orbit follows from Kepler's law. This is how the most precise black-hole masses outside the Local Group are measured, and how the megamaser distance ladder is built \citep{reid13,humphreys13,pesce20}. A disk that spans a factor of three or four in radius is, in principle, a direct test of whether the mass inside it is a point (Figure~\ref{fig:schematic}).

That test has a history that invites caution. The rotation curve of NGC~1068 is slower than Keplerian and was long interpreted as the signature of a massive, self-gravitating disk \citep{greenhill96_ngc1068,hure02,lodatobertin03}; more recent three-dimensional modelling suggests instead that the masers preferentially sample spiral arms away from the disk midline, with rotation consistent with Keplerian \citep{gallimore23}. Rotation-curve fits to seven megamaser disks pointed to substantial disk masses in some of them \citep{hure11}, whereas three-dimensional modelling of the same data found disk masses of order one per cent of the black hole or less \citep{kuo18}. In NGC~4258, a two-sigma, 0.8 per cent flattening of the rotation curve could be accounted for by an inclination warp of the disk \citep{herrnstein05}. In these cases an apparent departure from Keplerian rotation was first read as extended mass, and later found to be at least partly explained by the geometry of the disk.

Here we search for spikes in eleven megamaser disks with public maser catalogues. We fit the positions of the masers on the sky, in both coordinates, from their velocities, so that a bend in the maser line is measured rather than projected away; we let the disk warp, in position angle and in inclination, and give the warp the same freedom in the model with a spike and the model without one; and we calibrate every statistic and measure the sensitivity of every disk by injecting synthetic spikes into the real data. The result is a population statement: which disks could have shown a spike, which ones show a hint, and what observation would settle it.

\section*{Results}

\subsection*{Sample and model}

\begin{figure*}[p]
\centering
\includegraphics[width=0.96\textwidth]{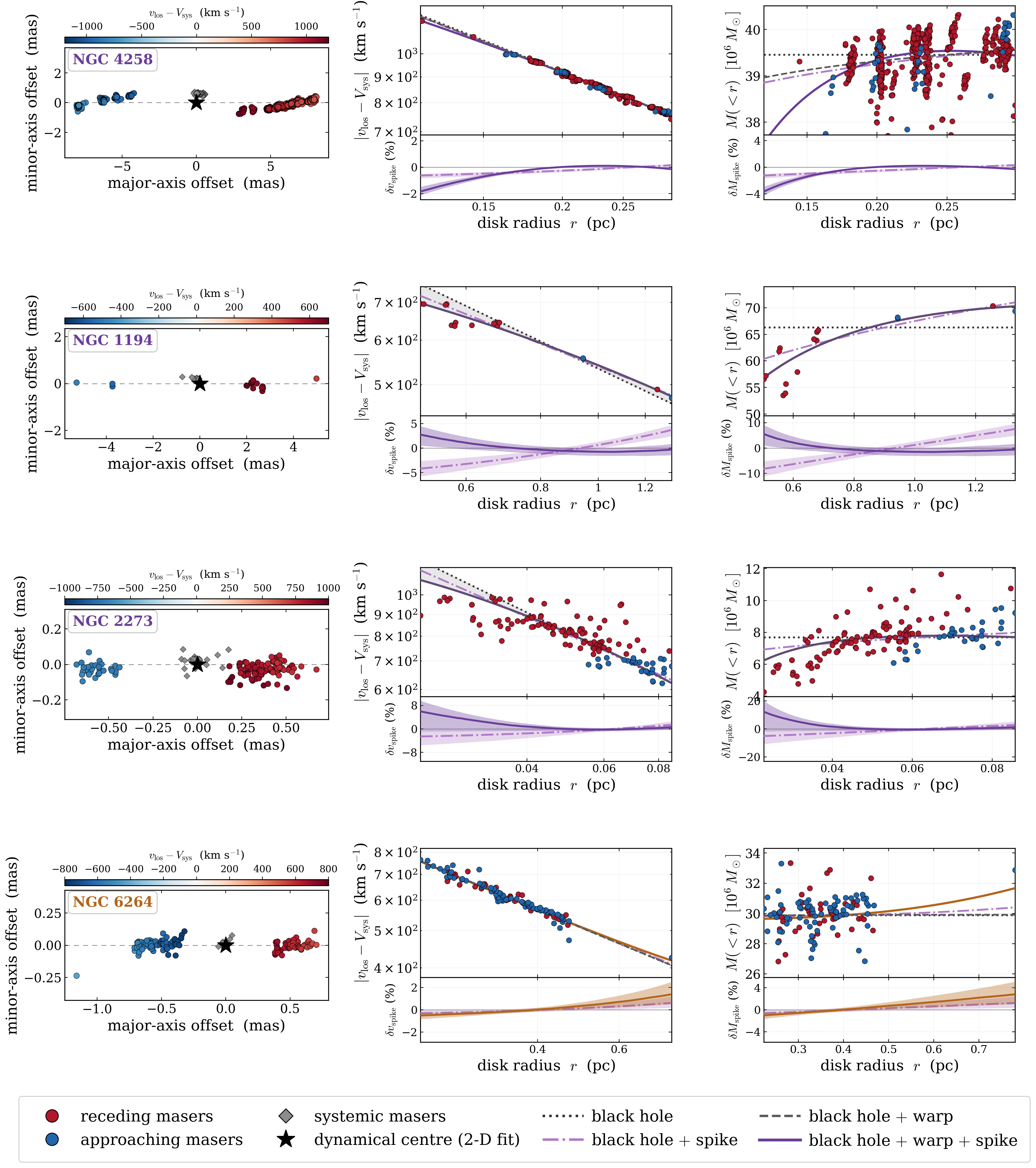}
\caption{Four disks: two positive-$\Delta$BIC candidates, NGC~4258 and NGC~1194; NGC~2273, whose enclosed mass rises most clearly; and NGC~6264 (orange), the one disk besides NGC~4258 in which a spike still improves the fit once the disk is warped. Left: maser positions in the disk frame, coloured by line-of-sight velocity (diamonds: systemic masers; star: fitted black hole). Centre: rotation curve with four models, black hole (dotted), black hole $+$ spike (dash-dotted), black hole $+$ warp (dashed) and black hole $+$ warp $+$ spike (solid). Right: enclosed mass $M(<r)=v^2r/G$ with the same models. Lower strips: median and $68\%$ band of the posterior spike models relative to the best fit without a spike, unwarped (light) and warped (dark), in velocity and in mass. Without a warp the rise is taken by a spike; with a warp the spike adds nothing in NGC~1194 and NGC~2273, while NGC~4258 and NGC~6264 keep a spike-like component (Methods).\label{fig:pv}}
\end{figure*}

\begin{figure*}[p]
\refstepcounter{table}\label{tab:main}
\noindent{\small\textbf{Table~\thetable.} Spike and warp in the eleven disks. $N$: high-velocity channels. The warp has two parts, a bend of the disk on the sky across the masing annulus ($\Delta{\rm PA}$) and a change of inclination along the line of sight ($\Delta i$). The next columns compare models with the same number of parameters: a spike added to the black hole in an unwarped disk, and a warp added to the same black hole, each by $\dlnl$ and $\Delta{\rm BIC}$ (positive favours the addition); then the gain from adding a spike to the warped disk. $\chi^2_{10\%}$: part of a noiseless ten-per-cent spike that the warp cannot absorb. Last column: 95th percentile of $\dM/\mbh$ (Figure~\ref{fig:contours}); parentheses mark values below the identifiability of the disk.}
\begin{center}\small
\setlength{\tabcolsep}{4pt}
\begin{tabular}{lrcc|cc|cc|ccc}
\hline\hline
 & & \multicolumn{2}{c|}{warp (deg)} & \multicolumn{2}{c|}{spike vs black hole} & \multicolumn{2}{c|}{warp vs black hole} & spike added & & \\
Galaxy & $N$ & $\Delta{\rm PA}$ & $\Delta i$ & $\dlnl$ & $\Delta{\rm BIC}$ & $\dlnl$ & $\Delta{\rm BIC}$ & to warp, $\dlnl$ & $\chi^2_{10\%}$ & $\dM/\mbh$ 95\% \\
\hline
NGC~5765b & 152 & $+8\pm2$ & $-8$ & $+7.5$ & $-2.5$ & $>100$ & $>100$ & $0.0$ & $14.4$ & (0.01) \\
NGC~2273 & 126 & $+29\pm4$ & $+32$ & $+2.0$ & $-7.7$ & $+32.9$ & $+23.3$ & $0.0$ & $1.0$ & $0.12$ \\
NGC~2960 & 125 & $+24\pm2$ & $+1$ & $0.0$ & $-9.7$ & $+43.6$ & $+33.9$ & $0.0$ & $1.7$ & (0.03) \\
NGC~6264 & 121 & $-14\pm2$ & $+3$ & $+0.5$ & $-9.1$ & $+57.1$ & $+47.5$ & $+6.5$ & $6.0$ & $0.07$ \\
CGCG~074-064 & 120 & $-16\pm2$ & $0$ & $0.0$ & $-9.6$ & $+26.9$ & $+17.3$ & $0.0$ & $16.5$ & (0.01) \\
NGC~6323 & 99 & $-4\pm2$ & $+2$ & $0.0$ & $-9.2$ & $+50.4$ & $+41.2$ & $0.0$ & $16.6$ & (0.01) \\
J0437+2456 & 66 & $-6\pm2$ & $-4$ & $+0.9$ & $-7.5$ & $+4.3$ & $-4.1$ & $0.0$ & $4.7$ & $0.12$ \\
ESO~558-G009 & 37 & $+6\pm10$ & $+16$ & $+2.6$ & $-4.6$ & $+5.2$ & $-2.0$ & $+0.1$ & $2.2$ & $0.31$ \\
NGC~1194 & 16 & $+7\pm1$ & $-19$ & $+11.5$ & $+6.0$ & $+16.2$ & $+10.7$ & $0.0$ & $5.9$ & $0.24$ \\
NGC~5495 & 8 & $+29\pm10$ & $+18$ & $+0.5$ & $-3.6$ & $+6.8$ & $+2.6$ & $+0.1$ & $0.0$ & $0.56$ \\
NGC~4258 & 600 & $-13\pm2$ & $-3$ & $+34.5$ & $+21.7$ & $>100$ & $>100$ & $+28.7$ & $>100$ & $0.05$ \\
\hline
\end{tabular}
\end{center}
\vspace{0.5ex}
\centering
\includegraphics[width=0.87\textwidth]{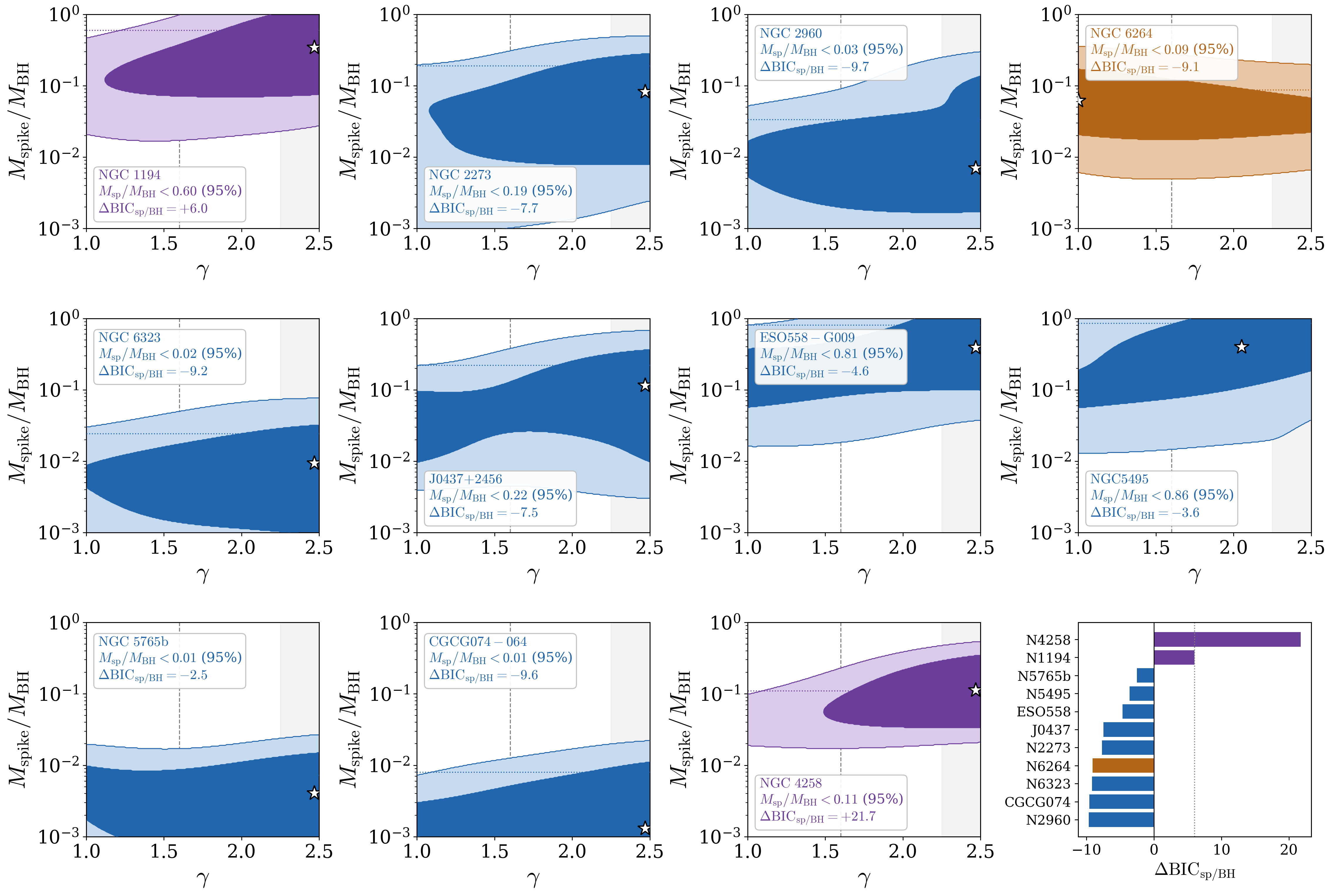}
\caption{Constraints on the spike in the eleven disks: $68\%$ and $95\%$ regions of spike slope $\gamma$ and spike mass $\msp/\mbh$, with the warp and the black-hole position marginalised (flat priors $\msp/\mbh<1$, $1\le\gamma\le2.5$). Star: posterior maximum; dotted line: 95th percentile of $\msp/\mbh$; grey band: adiabatic spikes, $\gamma=2.25$--$2.5$; dashed line: $\gamma\approx1.6$ \citep{sharma25}. Labels and the lower-right panel give $\Delta{\rm BIC}$ of a spike against a point mass in an unwarped disk, positive in two disks, shown in purple; orange marks NGC~6264, whose spike preference survives the warp.\label{fig:contours}}
\end{figure*}

The sample is listed in Extended Data Table~\ref{tab:sample}: eleven disks from five VLBI catalogues \citep{kuo11,gao16,gao17,pesce20,argon07}, with between 8 and 600 high-velocity maser channels each. Every disk is fitted with the same model. The black hole has mass $\mbh$ and may be surrounded by a spike of density slope $\gamma$ whose mass inside the outer maser radius is a fraction $f_{\rm sp}$ of the black hole's; the disk may bend on the sky by an angle $\Delta{\rm PA}$ across the masing annulus and tilt by an angle $\Delta i$ toward or away from edge-on; the position of the black hole is free within 0.2~milliarcseconds of the systemic masers. What a maser annulus measures directly is the increase of enclosed mass across it, $\dM/\mbh=f_{\rm sp}[1-(r_{\rm in}/r_{\rm out})^{3-\gamma}]$, and we quote that alongside $f_{\rm sp}$. The point-mass fits recover the published black-hole masses to within 8 per cent in every disk, and every disk needs a bend, from $4^\circ$ in NGC~6323 to $29^\circ$ in NGC~2273 (Table~\ref{tab:main} and Extended Data Table~\ref{tab:sample}; Extended Data Fig.~\ref{fig:bends}). Read directly from the masers, the enclosed mass $M(<r)=v^2r/G$ rises steadily with radius in three disks, NGC~2273, ESO~558-G009 and NGC~1194, with logarithmic slopes of $0.22$--$0.27$ at $5$--$8\sigma$ that survive the removal of any single maser; it is flat to within a few per cent in the others. Figure~\ref{fig:pv} shows two of them, NGC~1194 and NGC~2273, together with NGC~4258 and NGC~6264.

We compare a spike and a warp added to the black hole, each at the same cost in parameters, and then both together (Table~\ref{tab:main}). The spike slope is unconstrained in every disk: a maser annulus measures how much mass is added across it, not how that mass is distributed.

\subsection*{Spike versus point mass}

Against a black hole in an unwarped disk, a spike improves the fit in five disks, NGC~4258 ($\dlnl=34.5$), NGC~1194 ($11.5$), NGC~5765b ($7.5$), ESO~558-G009 ($2.6$) and NGC~2273 ($2.0$), and clears the BIC penalty in two of them, NGC~4258 ($\Delta{\rm BIC}=+21.7$) and NGC~1194 ($+6.0$; Table~\ref{tab:main} and Figure~\ref{fig:contours}). In NGC~1194, ESO~558-G009 and NGC~2273 the spike takes up the steady rise of the enclosed mass seen directly in the masers (Figure~\ref{fig:pv}, dash-dotted curves for NGC~1194 and NGC~2273). These spikes are too heavy to be the gas disk itself: a disk stable against its own gravity can weigh at most 0.5--1.3 per cent of the black hole (Methods), whereas the unwarped fits want 3--7 per cent of it in NGC~4258 and about 30 per cent in NGC~1194. They also lighten the black hole (Figure~\ref{fig:warpspike}, right): in NGC~1194, ESO~558-G009, NGC~2273, J0437+2456 and NGC~4258 the fitted mass drops to 0.58--0.81 of the published value while $\mbh(1+f_{\rm sp})$ stays within ten per cent of it, so the spike redistributes the enclosed mass rather than adding to it. A genuine extended component would do the same, because the published masses assume a point mass \citep{kuo11,gao16,gao17,pesce20}, and so would an unmodelled tilt of the disk; the lighter black hole alone does not tell the two apart, and the comparison with the warp below is what decides between them.

\subsection*{Warp versus point mass}

A spike and a warp do the same thing to a rotation curve. The line-of-sight velocity of a maser is $\sqrt{GM(<r)/r}\,\sin i$, so extra mass at large radius and a disk that turns more edge-on at large radius both make the outer masers move faster than Kepler predicts. The sky positions break part of this degeneracy: a warp bends the line of masers on the sky, and a spike does not. Every disk in the sample is bent (Extended Data Fig.~\ref{fig:bends}). The inclination part of the warp is not visible on the sky, and it is this part that competes with the spike. A warp, which costs the same two parameters as the spike, improves the fit of the black hole more than the spike does in all eleven disks, by $\dlnl=2.7$ to more than $100$ beyond the spike's gain (Table~\ref{tab:main} and Figure~\ref{fig:warpspike}, left). In NGC~1194, ESO~558-G009 and NGC~2273 the change of inclination alone reproduces the rise of the enclosed mass, and it does so without touching the black-hole mass (Figure~\ref{fig:pv}, dashed curves for NGC~1194 and NGC~2273).

\subsection*{Spike and warp combined}

\begin{figure*}[t]
\centering
\includegraphics[width=\textwidth]{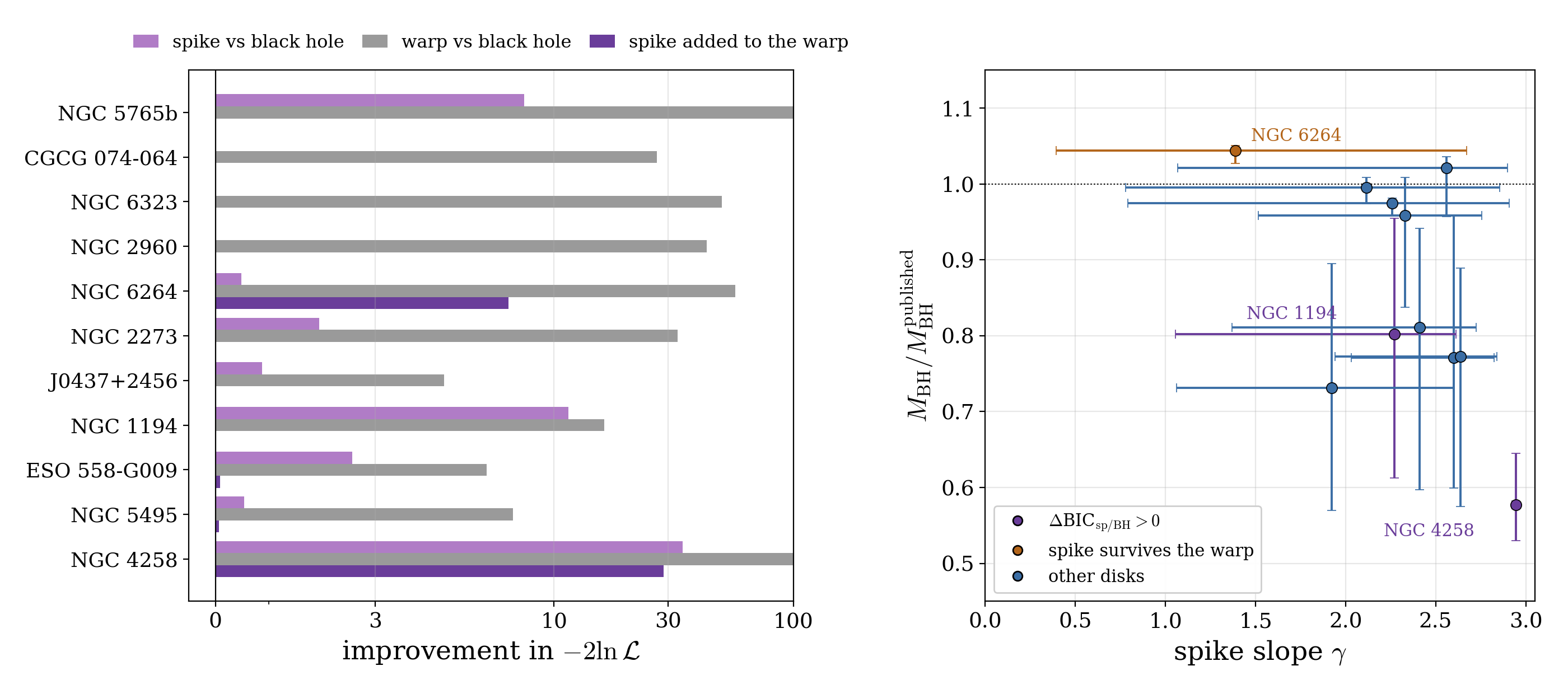}
\caption{Black hole alone versus a warp and a spike. Left: improvement in $-2\ln\mathcal{L}$ when a spike (light purple) or a warp (grey), each with two free parameters, is added to a black hole in an unwarped disk, and when a spike is added to the warped disk (dark purple). The warp fits better than the spike in all eleven disks; once it is in place the spike helps only in NGC~6264 and NGC~4258. Right: the spikes found when the disks are assumed unwarped: posterior median and $68\%$ interval of the spike slope and of the black-hole mass relative to the published value, purple, and labelled, for two disks with $\Delta{\rm BIC}>0$ in the unwarped fit (Figure~\ref{fig:contours}), and orange for NGC~6264, whose spike preference survives the warp. Every disk with a sizeable spike in the unwarped fit has its black hole lightened so that $\mbh(1+f_{\rm sp})$ stays at the published mass: the spike redistributes the enclosed mass rather than adding to it, as either a genuine extended component or a warp fitted as a spike would, and the comparison on the left, not the lighter black hole, discriminates between them; NGC~4258, whose warp is the best measured in the sample \citep{herrnstein05,humphreys13}, is the extreme case.\label{fig:warpspike}}
\end{figure*}

Once the disk is warped, adding a spike improves the fit in only two disks, NGC~6264 by $\dlnl=6.5$ and NGC~4258 by $28.7$ (Table~\ref{tab:main} and Figure~\ref{fig:warpspike}, left). In NGC~6264 the added mass sits mostly at the outer edge of the disk and is carried by the outermost masers (Figure~\ref{fig:pv}, orange). In the other nine it moves the rotation curve by no more than the scatter of the masers, and in NGC~1194 and NGC~2273 not at all (Figure~\ref{fig:pv}, lower strips). With the warp marginalised, the posteriors of Figure~\ref{fig:contours} bound the spike in eight disks; the 95 per cent upper limits on the mass added across the annulus range from about one per cent in best-sampled disks, to 12 per cent in NGC~2273 and J0437+2456; in three sparse disks, NGC~1194, ESO~558-G009 and NGC~5495, they reach 24--56 $\%$ and are set by the prior.

\subsection*{Joint constraints from the sample}

\begin{figure*}[t]
\centering
\includegraphics[width=\textwidth]{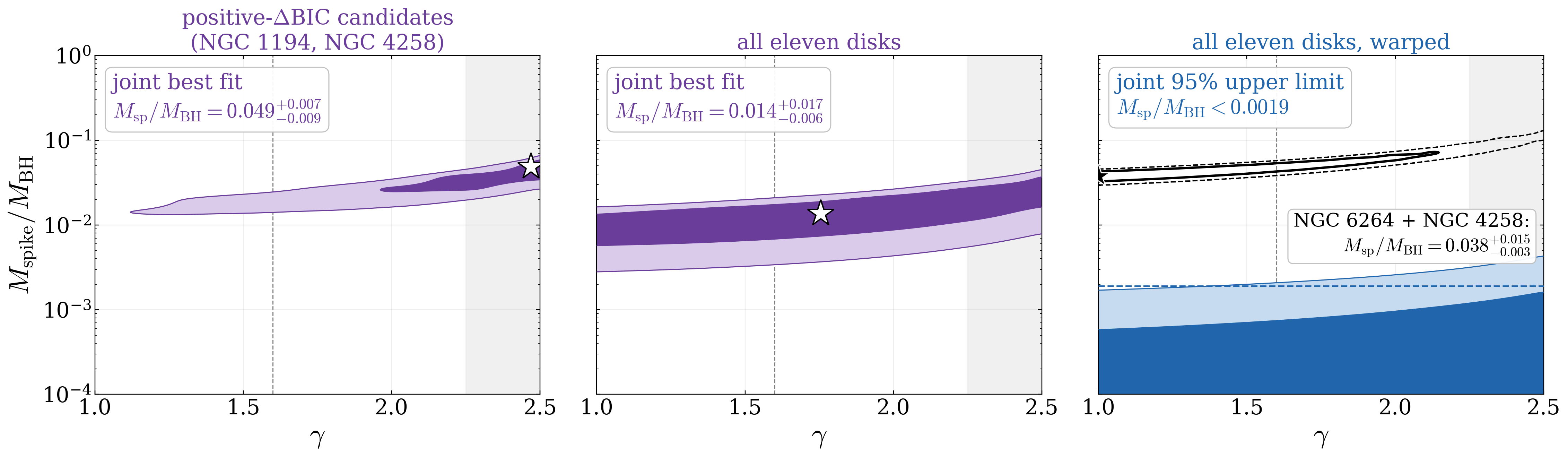}
\caption{Joint fit of a common spike, with the same slope $\gamma$ and mass fraction $\msp/\mbh$ in every disk. Regions enclose $\Delta\chi^2<2.30$ and $6.18$ (68-95 $\%$ for two parameters) of the summed profile likelihood, each disk's black-hole mass, geometry, centre and scatter being refitted at every grid point. Left: NGC~1194 and NGC~4258, with $\Delta{\rm BIC}>0$ against a black hole alone, in unwarped disks. Centre: all eleven disks, unwarped. Right: all eleven disks, warped; the dashed line is the 95 per cent upper limit on $\msp/\mbh$ for the least constrained slope, and the black contours are the joint fit of NGC~6264 and NGC~4258 alone, whose spike preference survives the warp (on a finer grid of 13 slopes and 30 mass fractions). Stars mark the joint best fits; grey band and dashed vertical line as in Figure~\ref{fig:contours}.\label{fig:joint}}
\end{figure*}

Figure~\ref{fig:joint} combines the disks under the assumption that every black hole carries a spike with the same slope and the same mass fraction, by adding the profile likelihoods of the disks on a common grid of $\gamma$ and $\msp/\mbh$ (Methods). Without a warp, NGC~1194 and NGC~4258, which clear the BIC penalty, together prefer $\msp/\mbh=0.049^{+0.007}_{-0.009}$ at the steep end of the slope prior, $\Delta\chi^2=34$ better than no spike, although NGC~1194 alone wants a heavier spike and pays $\Delta\chi^2\approx10$ at the common solution. Adding the other nine disks lowers the joint best fit to $\msp/\mbh=0.014^{+0.017}_{-0.006}$, still $\Delta\chi^2=11.7$ better than no spike but in tension with individual disks: CGCG~074-064, NGC~1194, NGC~4258 and NGC~5765b each pay $\Delta\chi^2=6$--$12$ at the common solution. Once the disks may warp, the combination prefers no spike and gives a 95 per cent upper limit $\msp/\mbh<0.002$ for any slope in the prior range, from $6\times10^{-4}$ at $\gamma=1$ to $0.002$ at $\gamma=2.5$, set mainly by NGC~5765b and CGCG~074-064; NGC~4258 is in tension with this limit by $\Delta\chi^2=29$, the residual preference discussed below. NGC~6264 and NGC~4258, which keep a spike once they are warped, combine into a narrow band at $\msp/\mbh=0.038^{+0.015}_{-0.003}$, $\Delta\chi^2=31$ better than no spike, with its best fit at the shallow edge of the prior (black contours in Figure~\ref{fig:joint}); it lies an order of magnitude above the sample limit and is set by two disks whose geometry the linear warp describes least well (below).

\subsection*{Sensitivity of the search}

Because the warp absorbs part of any spike, we measured the sensitivity of the search with the warp free, in two steps (Extended Data Fig.~\ref{fig:sensitivity}). First, without noise: a spike of $\gamma=2.25$ adding 3, 10 or 30 per cent of the black-hole mass across the annulus was added to each disk's best-fitting warped point mass, exact maser positions were generated at the observed velocities, and the warped point mass was refitted. Whatever $\chi^2$ is left is the part of the spike that geometry cannot hide. The warp hides 97 to 99.9 per cent of the spike response in every disk. A three-per-cent spike is invisible everywhere. A ten-per-cent spike leaves about $4\sigma$ in three best-sampled disks of the Megamaser Cosmology Project, NGC~5765b, CGCG~074-064 and NGC~6323, $2.5\sigma$ in NGC~6264 and NGC~1194, and only $1\sigma$ in NGC~2273. In NGC~4258, whose 600 masers are the most precise in the sample, it leaves $\chi^2>100$, although the warp still hides 99.9 per cent of the response.

Second, with noise: for four disks with power we generated 100 spike-free mock disks and 60 mocks with spikes of 5, 10 and 20 per cent on each disk's own sampling and errors, refitted both models on every mock with the identical procedure, and asked how often the spike was detected above the 95th percentile of the spike-free distribution. A ten-per-cent spike is detected every time in NGC~5765b, CGCG~074-064 and NGC~6323, and in $83\pm10$ per cent of realisations in NGC~6264; a five-per-cent spike in 50 to 80 per cent (Extended Data Table~\ref{tab:power}). Against that, the real data give no improvement at all in NGC~5765b, CGCG~074-064 and NGC~6323. A spike of ten per cent of the black-hole mass across the annulus is excluded in those three disks with 90--100 $\%$ confidence, and one of five per cent with 50--80 $\%$ confidence.

\subsection*{Residual spike preference in NGC~6264 and NGC~4258}

NGC~6264 is the disk with a spike preference that survives the warp, at $\dlnl=6.5$. Among the 100 spike-free mock disks, two exceed that value, so the local significance is about $2\sigma$; the preferred component is shallow, with $\gamma$ at the lower edge of the prior, and 31 of the 100 spike-free mocks find the same shape, so it is not a rare one. The added mass is about seven per cent of the black hole, below the ten per cent at which the recovery is reliable. NGC~4258 prefers a spike much more strongly, at $\dlnl=28.7$, with $f_{\rm sp}=0.66$ at $\gamma=2.88$ and the black hole at $0.59$ of its published mass, the same redistribution seen in the unwarped fits. It is also the one disk whose geometry has been measured in detail: its warp needs a curved position angle and a change of inclination to describe \citep{herrnstein05,humphreys13}, richer than the linear warp used here, and three-dimensional modelling puts the mass of its disk at about 0.1 per cent of the black hole \citep{kuo18}. We therefore read both cases the same way: as disks whose geometry the linear warp describes least well, and in which a spike and a more complex warp remain to be separated.

\section*{Discussion}

We showed that without a warped disk, several megamaser rotation curves prefer extra mass. A spike improves the fit over a black hole alone in five disks and clears the BIC penalty in NGC~1194 and NGC~4258, which together favour a spike of about five per cent of the black-hole mass. However, a single rotation curve observed at a single epoch can hardly, on its own, tell such a spike from a warp; a change of inclination of $10$ to $15$ degrees across the annulus reproduces the velocity signature of a spike of several tenths of the black-hole mass, and the bend that a warp leaves on the sky does not fix its inclination part. Yet even when the disks are allowed to warp, NGC~4258 and NGC~6264 retain a preference for a spike, jointly at about four per cent of the black-hole mass (Figure~\ref{fig:joint}). This sits above the limit from the full sample, pointing either to spikes that differ from galaxy to galaxy or to geometry richer than a linear warp, and it calls for dedicated follow-up. What this work also shows is that the degeneracy can be measured disk by disk: in the best-sampled disks a spike of ten per cent of the black-hole mass would have been seen at $4\sigma$ even with the warp free, and the eleven disks together limit a common spike to at most 0.2 per cent of the black-hole mass. These are the first geometric, population-level constraints on dark matter spikes around supermassive black holes, and they come from public data alone.

The measurement that would break the degeneracy is already being made. The Megamaser Cosmology Project monitors its disks for distances \citep{reid13,humphreys13,gao16,pesce20}, and the line-of-sight accelerations it records do two things a rotation curve cannot. Accelerations of the high-velocity masers fix their azimuth, removing the off-midline effect that misled the NGC~1068 analyses, a galaxy whose neutrino and electromagnetic emission is also used to probe the particle nature of dark matter \citep{cline23,herrera25,herreramurase24}. Accelerations of the systemic masers weigh the black hole at a second, smaller radius: a spike predicts less mass there than at the outer edge of the disk, a warp predicts the same. Two to three years of VLBI monitoring of NGC~6264 and NGC~1194 would deliver both, and would map each warp independently of its rotation curve, deciding whether the excesses that five disks show when assumed unwarped (Table~\ref{tab:main}) are spikes or geometry. The ngVLA \citep{murphy18} and the SKA \citep{braun15} will then find fainter masers in the inner annuli, where a spike and a tilt differ most, and many more disks. Megamaser disks can now ask, black hole by black hole and across a population, whether dark matter gathers where gravity is strongest.

\section*{Methods}

\subsection*{Data and sample}

We compiled every megamaser-disk galaxy for which per-feature VLBI kinematics, sky-position offsets and line-of-sight velocities of individual maser features, are publicly available in machine-readable form (Extended Data Table~\ref{tab:sample}). The five catalogues use heterogeneous conventions: \citet{kuo11} and \citet{gao17} list maser offsets in equatorial coordinates relative to a reference position; \citet{gao16} and \citet{pesce20} use Cartesian $(X,Y)$ offsets in a frame aligned with the disk; \citet{argon07} provides per-channel detections across 18 epochs. All five rest on $22$~GHz VLBI observations with the Very Long Baseline Array (VLBA), in most cases augmented by one or more of the 100-m Green Bank Telescope (GBT), the Effelsberg 100-m telescope and the phased Karl G. Jansky Very Large Array (VLA). \citet{kuo11} observed with the VLBA and the GBT, and in most cases Effelsberg; \citet{gao16} with the VLBA, GBT, Effelsberg and VLA between 2012 April and 2014 January; \citet{gao17} with the VLBA and GBT between 2010 and 2014, some tracks adding Effelsberg and the phased VLA; \citet{pesce20} with the VLBA alone for the phase-referenced track and with the High Sensitivity Array (VLBA, GBT and phased VLA) for the self-calibrated tracks; and \citet{argon07} over eighteen epochs between 1997 and 2000, twelve with the VLBA alone and six with the VLBA, the phased VLA and Effelsberg. All five catalogues tabulate a position error in each coordinate for every entry, which the likelihood uses. A single loader ingests the five formats and, for each galaxy, isolates the high-velocity channels (those with $|v-v_{\rm sys}|$ above a per-catalogue threshold of 150, 250 or 500$\kms$, or with a flagged spot-type membership where the catalogue provides one) and the systemic channels, whose two-dimensional centroid defines the reference position for the dynamical centre. The systemic velocity is adopted from the discovery paper where tabulated, from the systemic-maser cluster mean where the catalogue provides a feature type, and from a least-scatter Keplerian fit otherwise. The outer inclination $i_0$ is taken from the discovery paper of each disk; the fitted mass absorbs $\sin^2 i_0$, and only the change of inclination across the annulus is fitted. For NGC~4258 the 600 channels are a fixed random subset of the high-velocity entries across all epochs, with epoch identity discarded; its geometry is known to require a curved position-angle warp and an inclination warp \citep{humphreys13}, richer than the linear warp fitted to every disk here, and its results should be read with that in mind. Throughout we use $M(<r)=v^2r/G=232.4\,v^2 r$, with $M$ in $\msun$, $v$ in $\kms$ and $r$ in parsecs.

\subsection*{Treatment of catalogue entries}

The entries of these catalogues are elliptical-Gaussian fits to individual spectral-channel maps \citep{kuo11}: in NGC~2273, 87 per cent of consecutive high-velocity entries are exactly one channel width ($1.72\kms$) apart, and the 126 entries form 17 velocity-contiguous runs. The intraclass correlation of the residuals of channels within a run is $0.04$ in velocity and $0.14$ in position, so the channels are close to independent measurements of the position of their feature at a set of exactly known velocities, and that is how they enter the likelihood. Treating them instead as independent measurements of velocity at exactly known positions inverts the roles of the well-measured and the noisy quantity: the position of an inner NGC~2273 maser at $0.2$--$0.3$~mas carries a catalogue error of $0.03$--$0.04$~mas, so the radius as an independent variable carries 15--20 per cent noise, and ordinary regression of $\ln v$ on $\ln r$ then returns a slope attenuated toward zero, $-0.4$ rather than $-0.5$, which reads as an enclosed mass rising outward. \citet{kuo11} fit position against velocity for this reason, and \citet{kuo18} showed that the projected position--velocity method returns spurious extended mass of either sign in Keplerian disks.

\subsection*{Two-dimensional disk model}

For parameters $\theta$, the rotation curve on a radial grid is
\begin{equation}
v(r)=\sqrt{\frac{G\mbh\,[1+f_{\rm sp}(r/r_{\rm out})^{3-\gamma}]}{r}}\;\frac{\sin[i_0+\Delta i\,h(r)]}{\sin i_0},
\label{eq:vr}
\end{equation}
with $h(r)=(\ln r_{\rm out}-\ln r)/(\ln r_{\rm out}-\ln r_{\rm in})$ clipped to $[0,1]$, so that $h=1$ at the inner and $0$ at the outer edge of the masing annulus, and $r_{\rm in}$, $r_{\rm out}$ fixed for each disk from the extreme high-velocity radii about the systemic centroid. The curve is inverted for $r(v)$ on the monotonic branch that contains the observed annulus; a model in which some observed velocity has no root on that branch is rejected rather than extrapolated. A maser on the midline at radius $r$ on side $s=\pm1$ of the disk, where the ring plane has position angle $\Omega(r)=\Omega_0+\Delta\Omega\,h(r)$ east of north, is predicted at
\begin{equation}
X=x_0+s\,\frac{r\sin\Omega(r)}{D},\qquad Y=y_0+s\,\frac{r\cos\Omega(r)}{D},
\end{equation}
with $D$ the distance in parsecs per milliarcsecond. The likelihood is Gaussian in $X$ and $Y$ with variances $e_{X,i}^2+s_X^2$ and $e_{Y,i}^2+s_Y^2$, where $e$ are the catalogue errors and $s_X$, $s_Y$ fitted excess scatters, plus the Gaussian centre prior $[(x_0-x_{\rm sys})^2+(y_0-y_{\rm sys})^2]/(0.2\,{\rm mas})^2$. The angular parameters $\Delta\Omega$ and $\Delta i$ are bounded at $\pm45^\circ$ in every fit; $\Omega_0$ within $\pm0.5$~rad of the receding side's mean position angle. Six nested models are fitted: a point mass (six parameters: $\mbh$, $\Omega_0$, $x_0$, $y_0$, $s_X$, $s_Y$), with a bend ($+\Delta\Omega$), with a bend and an inclination change ($+\Delta i$), and each of these with a spike ($+f_{\rm sp}$, $\gamma$, with $0\le f_{\rm sp}\le1$ and $0.05<\gamma<2.95$). Each is maximised by differential evolution from two or three seeds followed by a Nelder--Mead polish and warm starts from the nested models; the spike model is never allowed to be worse than the model it contains. No term for the mass of the gas disk is included; the section on disk self-gravity below shows why.

\subsection*{Centre prior}

The systemic masers lie on the near side of the disk and their centroid need not coincide with the black hole to better than the asymmetry of their arc. J0437+2456 makes this quantitative: with the centre pinned to the statistical error of its systemic centroid ($0.003$ and $0.005$~mas), the receding and approaching halves give enclosed masses of $1.32$ and $0.70$ of the published value and the excess scatter is $0.021$~mas; with the centre free, both halves give $0.91$--$0.92$, the scatter drops to $0.005$~mas, $-2\ln\mathcal{L}$ improves by $290$, and the black hole sits $0.108$~mas from the systemic centroid along the disk. A prior of $0.2$~mas in both coordinates, the same for every source, accommodates that offset while excluding larger excursions of the centre, which by themselves can produce an apparent slope in the enclosed mass. Repeating the sample at $0.01$, $0.05$ and $0.4$~mas changes no conclusion of Table~\ref{tab:main}: NGC~2273 gives $f_{\rm sp}=0$ at every width, and NGC~1194's spike-versus-warp contest moves by less than one unit of $-2\ln\mathcal{L}$.

\subsection*{Statistics}

Table~\ref{tab:main} reports differences in $-2\ln\mathcal{L}$ between maximum-likelihood fits. Adding a spike or adding a warp to the black hole each costs two parameters, so the two additions are compared directly; each is also scored with the Bayesian information criterion, $\Delta{\rm BIC}=\dlnl-2\ln N$ with $N$ the number of channels \citep{schwarz78,kassraftery95}. By this criterion a spike beats the black hole alone in NGC~4258 and NGC~1194, and a warp beats it in nine of the eleven disks. Because the spike amplitude sits on a boundary of its prior under the null and the slope is unidentified there, we do not translate either statistic into a significance; instead the statistic $T$, the improvement in $-2\ln\mathcal{L}$ from adding the spike to the warped point mass, is calibrated on spike-free mocks generated on each disk's own sampling, as described next, and the significance quoted for NGC~6264 is the fraction of those mocks exceeding its observed $T$.

\subsection*{Identifiability and injection protocol}

For each disk, the best-fitting warped point mass is taken as the truth and a spike with $\gamma=2.25$ and annular increment $\dM/\mbh=\delta$ is added, which sets $f_{\rm sp}=\delta/[1-(r_{\rm in}/r_{\rm out})^{0.75}]$; maser positions are then generated at the observed channel velocities. In the noiseless test these exact positions are refitted with the warped point mass (mass, position angle, bend, inclination change and centre free; excess scatters fixed at their fitted values), and the $\chi^2$ of that fit relative to the truth is the part of the spike the warp cannot absorb (Extended Data Fig.~\ref{fig:sensitivity}, left). A linearised (Fisher) projection of the spike onto the nuisance parameters confirms that only 0.1--3.5 per cent of the spike response is orthogonal to them, although it overestimates the residual $\chi^2$ because the absorption is non-linear. In the noisy test, Gaussian noise with the catalogue errors and fitted excess scatter is added, both models are refitted from two starting points (the generating geometry and the real-data solution) with all nuisances free, and $T$ is recorded. We ran 100 spike-free and 60 injected mocks per amplitude for NGC~5765b, CGCG~074-064, NGC~6323 and NGC~6264. Departures from the generating model (offsets in inclination, off-midline azimuths, peculiar velocities, curved warps, correlated astrometry) are not included, so the power curves are conditional on the observed sampling and on the linear-warp model.

\subsection*{Posteriors}

The posteriors of Figure~\ref{fig:contours} sample the same likelihood with the affine-invariant stretch move of \citet{goodman10} in an independent \texttt{numpy} \citep{numpy} implementation, 40 walkers and 4000 steps with 1500 discarded, under flat priors $f_{\rm sp}\in[0,1]$, $\gamma\in[1,2.5]$, $|\Delta\Omega|,|\Delta i|\le45^\circ$ and the centre prior above. Split-half 95th percentiles of $f_{\rm sp}$ agree to a few per cent in every disk except NGC~2273 ($0.16$ and $0.24$). Widening the slope prior to $\gamma<2.95$ moves the 95th percentile of $f_{\rm sp}$ in NGC~2273 from $0.18$ to $0.78$ but that of $\dM/\mbh$ only from $0.11$ to $0.15$, which is why the increment and not the amplitude is the quantity we report. Where a posterior is narrower than the identifiability of the disk (the parenthesised entries of Table~\ref{tab:main}), the two numbers answer different questions. A tilt can mimic a spike, because turning a nearly edge-on disk away from edge-on toward its centre lowers the inner velocities, but it can raise them only until the disk is edge-on, by at most $1-\sin i_0$. A disk whose rotation curve shows no rise therefore bounds a spike more tightly than it could tell a spike from a tilt. The bound does not come from the $\pm45^\circ$ range of the angles: widening it to $\pm80^\circ$ leaves the profile likelihood of NGC~5765b unchanged. The unwarped-disk posteriors of Figure~\ref{fig:warpspike} (right) sample the spike model with $\Delta\Omega=\Delta i=0$, $\gamma\in[0.05,2.95]$, 32 walkers and 1500 steps.

\subsection*{Joint fit}
For the population fit of Figure~\ref{fig:joint} each disk was refitted on a grid of seven slopes ($\gamma=1$--$2.5$) and eighteen mass fractions ($\msp/\mbh=0$--$0.8$), minimising $-2\ln\mathcal{L}$ over its black-hole mass, position angle, centre and excess scatter, and in the warped case also over the bend and the tilt, from several starting points including the best fits with and without a spike. The grids of the disks were summed and interpolated with shape-preserving cubics in $\gamma$ and $\log(\msp/\mbh+5\times10^{-4})$. A common mass fraction is the simplest population hypothesis, a spike whose mass scales with that of its black hole. Because $\msp$ is measured inside each disk's own outer maser radius, which ranges from 0.09 to 1.3~pc across the sample, the common value is a common mass fraction within the region each disk samples rather than a common physical density profile; the tensions quoted in the text measure how well each disk tolerates it.

\subsection*{Disk self-gravity}
The spike term measures any mass distributed across the masing annulus, and the gas disk itself contributes such mass. A self-gravitating disk whose surface density falls as $\Sigma\propto r^{-p}$ encloses $M_{\rm d}(<r)=2\pi\Sigma(r)\,r^2/(2-p)$, which grows with radius much like a spike: $p=1$ gives $M_{\rm d}\propto r$, the behaviour of $\gamma=2$. How massive can the disk be? A thin disk around a point mass is stable against its own gravity only if the Toomre parameter $Q=c_s\kappa/(\pi G\Sigma)$ exceeds unity \citep{toomre64,goodman03}, where $c_s$ is the sound speed and the epicyclic frequency equals the orbital frequency, $\kappa=\Omega=v/r$, for Keplerian rotation. Setting $Q\ge1$ gives $\Sigma\le c_s v/(\pi G r)$ and hence
\begin{equation}
\frac{M_{\rm d}(<r)}{\mbh}\le\frac{2}{(2-p)\,Q}\,\frac{c_s}{v}\simeq\frac{2}{(2-p)\,Q}\,\frac{H}{r},
\label{eq:toomre}
\end{equation}
using $\mbh=v^2r/G$ and hydrostatic equilibrium, $H/r=c_s/v$: a stable disk can weigh no more than about its aspect ratio times the black-hole mass. Water masers are excited in warm, dense molecular gas at a few hundred kelvin and above \citep{lo05}; at $1000$~K, near the top of that range, molecular gas has $c_s\approx1.9$~km~s$^{-1}$, while the orbital speeds at the outer edges of the eleven disks are 300--740~km~s$^{-1}$. For $p=1$ and $Q=1$, equation~(\ref{eq:toomre}) therefore bounds the disk at 0.5-1.3 $\%$ of the black-hole mass, the smallest value for NGC~4258 (0.5 per cent) and NGC~2273 (0.6 per cent) and the largest for NGC~2960 and J0437+2456 (1.2--1.3 $\%$). This is generous: it takes the warmest masing gas and a marginally stable disk.

Most of what this paper reports lies well above that bound. The spikes preferred when the disks are assumed unwarped are heavier than any stable disk: NGC~1194 wants $\msp/\mbh\approx0.3$ against a bound of 0.8 per cent, NGC~4258 about 3--7 per cent against 0.5 per cent, and the joint fit of NGC~1194 and NGC~4258 $4.9$ per cent (Figure~\ref{fig:joint}), so none of these excesses can be the gravity of a stable gas disk. The joint fit of all eleven unwarped disks, $1.4^{+1.7}_{-0.6}$ per cent, is comparable to the bound and could not by itself be distinguished from a disk. The upper limits from the warped fits exceed the bound in nine disks and are comparable to it in NGC~5765b and CGCG~074-064 (0.7 and 0.8 per cent against 0.9 and 0.8 per cent); only the joint limit, $\msp/\mbh<0.2$ per cent, lies below it. There a disk would matter, but only in one direction: its mass adds to that of any spike, so a limit on the total extended mass is also a limit, at least as strict, on the dark matter. A disk heavier than equation~(\ref{eq:toomre}) allows would have to be gravitationally unstable and clumpy, as has been argued for some megamaser disks from their rotation curves \citep{hure11}; such a disk would raise, not lower, the mass available to mimic a spike in an unwarped fit, and it would be a further reason to read those fits as geometry and baryons rather than dark matter.

\subsection*{Limitations}

The warp is linear in $\ln r$ in both angles, the masers are assumed to lie on the midline, $i_0$ and the systemic velocity are fixed from the discovery papers, and peculiar velocities are not modelled. Each of these can move mass between the spike and the geometry in either direction. The upper limits in particular rest on the published outer inclinations: a tilt toward edge-on can raise the inner velocities by at most $1-\sin i_0$, which for these disks allows between 0 and 2.5 per cent of the black-hole mass to be cancelled, comparable in the best-constrained disks to the joint limit; the acceleration data of NGC~5765b and CGCG~074-064 bound the midline assumption to below two per cent in velocity for those two disks (mean $\sin\phi=-0.007\pm0.022$ and $-0.008\pm0.028$ for the inner high-velocity masers), and the inclination change is the parameter through which a spike is most easily absorbed, so freeing $i_0$ within its published uncertainty would widen, not narrow, the constraints. The model comparison is reported with the maximised likelihood including the centre prior, which enters both models identically. Natural extensions of this work are an epoch-resolved treatment of NGC~4258 with its published geometry, injection tests under departures from the generating model, and a joint fit of positions, velocities and accelerations.

\subsection*{Data availability}

All maser-spot positions, velocities and, where available, accelerations used in this work are taken from the published tables of \citet{kuo11}, \citet{gao16}, \citet{gao17}, \citet{pesce20} and \citet{argon07}. The fitted models, mock catalogues and posterior samples are available from the author on reasonable request.

\subsection*{Code availability}

The analysis code, written in Python with \texttt{numpy} \citep{numpy} and \texttt{matplotlib} \citep{matplotlib}, is available from the author on reasonable request.

\begin{acknowledgments}
We thank Mark Reid for useful feedback on this manuscript. This work was supported by the Neutrino Theory Network Fellowship with contract number 726844.
\end{acknowledgments}

\vspace{1ex}\noindent\textbf{Author contributions.} G.H. conceived the project, performed the analysis and wrote the manuscript. AI-based tools assisted with code development, parts of the analysis and editing of the text. The scientific question, the approach and the interpretation of the results are the author's own, and the author takes full responsibility for the content.

\vspace{1ex}\noindent\textbf{Competing interests.} The author declares no competing interests.

\clearpage
\section*{Extended Data}

\renewcommand{\thefigure}{ED\arabic{figure}}
\renewcommand{\thetable}{ED\arabic{table}}
\setcounter{figure}{0}
\setcounter{table}{0}

\onecolumngrid
\refstepcounter{table}\label{tab:sample}
\noindent{\small\textbf{Table~\thetable.} The eleven-galaxy megamaser-disk sample. $N$ is the number of high-velocity maser channels used; the last column is the black-hole mass recovered by the warped point-mass fit of this work, divided by the published value.}
\begin{center}\small
\begin{tabular}{lcrrrcc}
\hline\hline
Galaxy & VLBI source & $D$ & $\mbh^{\rm pub}$ & $N$ & $r$ range & $\mbh^{\rm kep}/\mbh^{\rm pub}$ \\
 & & (Mpc) & ($10^7\msun$) & & (pc) & \\
\hline
NGC~1194 & \citet{kuo11} & 52 & 6.40 & 16 & 0.55--1.30 & 1.04 \\
NGC~2273 & \citet{kuo11} & 26 & 0.76 & 126 & 0.02--0.09 & 1.01 \\
NGC~2960 & \citet{kuo11} & 71 & 1.14 & 125 & 0.09--0.55 & 1.00 \\
NGC~6264 & \citet{kuo11} & 136 & 2.84 & 121 & 0.21--0.77 & 1.05 \\
NGC~6323 & \citet{kuo11} & 105 & 0.93 & 99 & 0.12--0.30 & 1.04 \\
J0437+2456 & \citet{gao17} & 65.3 & 0.29 & 66 & 0.04--0.12 & 0.92 \\
ESO~558-G009& \citet{gao17} & 107.6 & 1.70 & 37 & 0.20--0.43 & 1.00 \\
NGC~5495 & \citet{gao17} & 95.7 & 1.10 & 8 & 0.11--0.37 & 0.93 \\
NGC~5765b & \citet{gao16} & 126.3 & 4.55 & 152 & 0.25--1.21 & 1.04 \\
CGCG~074-064& \citet{pesce20} & 87.6 & 2.42 & 120 & 0.12--0.47 & 0.98 \\
NGC~4258 & \citet{argon07} & 7.6 & 4.00 & 600 & 0.11--0.31 & 0.99 \\
\hline
\end{tabular}
\end{center}
\vspace{1ex}
\refstepcounter{table}\label{tab:power}
\noindent{\small\textbf{Table~\thetable.} Sensitivity of the four identifiable disks from noisy injection--recovery (Methods). $T^{\rm null}_{95}$ is the 95th percentile of the likelihood-ratio statistic over 100 spike-free mocks; the power columns give the fraction of 60 injected mocks exceeding it at each annular increment, with the median recovered increment in parentheses; the last column is the statistic observed in the real data and the fraction of null mocks exceeding it.}
\begin{center}\small
\begin{tabular}{lcccc}
\hline\hline
Galaxy & $T^{\rm null}_{95}$ & power at $\dM/\mbh=0.05$ & $0.10$ & observed $T$ ($p_{\rm null}$) \\
\hline
NGC~5765b    & 0.56 & 0.82 (0.03) & 1.00 (0.07) & 0.0 (0.41) \\
CGCG~074-064 & 2.22 & 0.53 (0.03) & 1.00 (0.08) & 0.0 ($>0.5$) \\
NGC~6323     & 0.79 & 0.75 (0.03) & 1.00 (0.07) & 0.0 ($>0.5$) \\
NGC~6264     & 2.92 & 0.30 (0.04) & 0.83 (0.07) & 6.5 (0.03) \\
\hline
\end{tabular}
\end{center}
\twocolumngrid

\begin{figure*}
\centering
\includegraphics[width=\textwidth]{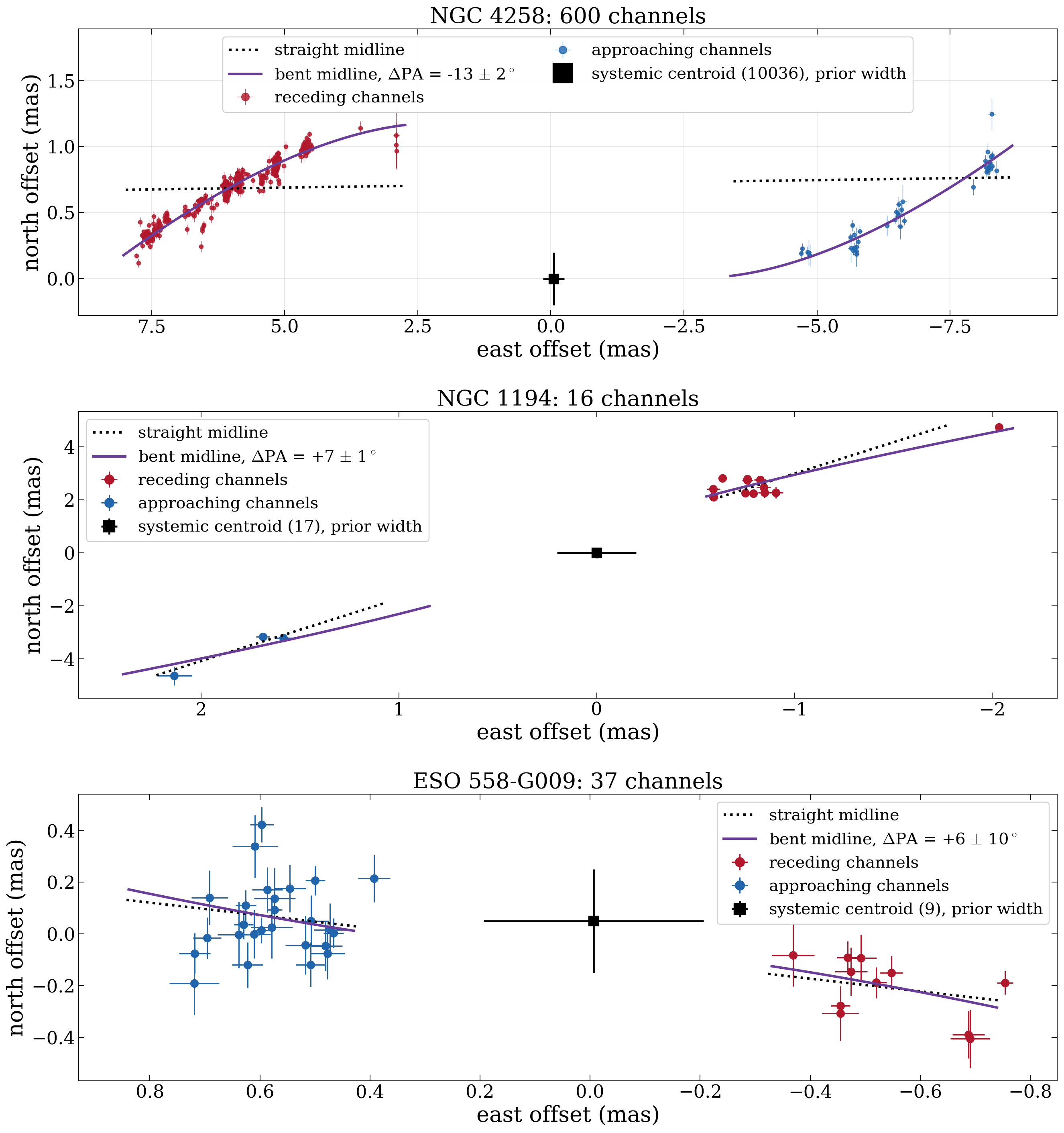}
\caption{Positions of the maser channels in the two candidates, NGC~4258 (top, data from \citealt{argon07}) and NGC~1194 (middle, \citealt{kuo11}), and in ESO~558-G009 (bottom, \citealt{gao17}), with the north axis stretched relative to the east axis to show the bends; receding in red and approaching in blue with their catalogue errors; the black square is the centroid of the systemic masers with the $0.2$~mas prior width used for the position of the black hole. Lines are the midline of the disk fitted in two dimensions to all channels with a point mass: straight (dotted) and with a position-angle bend that varies linearly with $\ln r$ across the masing annulus (solid), which improves $-2\ln\mathcal{L}$ by more than $100$, by $5$ and by $0.3$ respectively. The bend is the visible half of a warp; its invisible half, the change of inclination with radius, multiplies the line-of-sight velocities by $\sin i(r)$ and is degenerate with a rising enclosed mass in any one-dimensional rotation curve.\label{fig:bends}}
\end{figure*}

\begin{figure*}[h!]
\centering
\includegraphics[width=\textwidth]{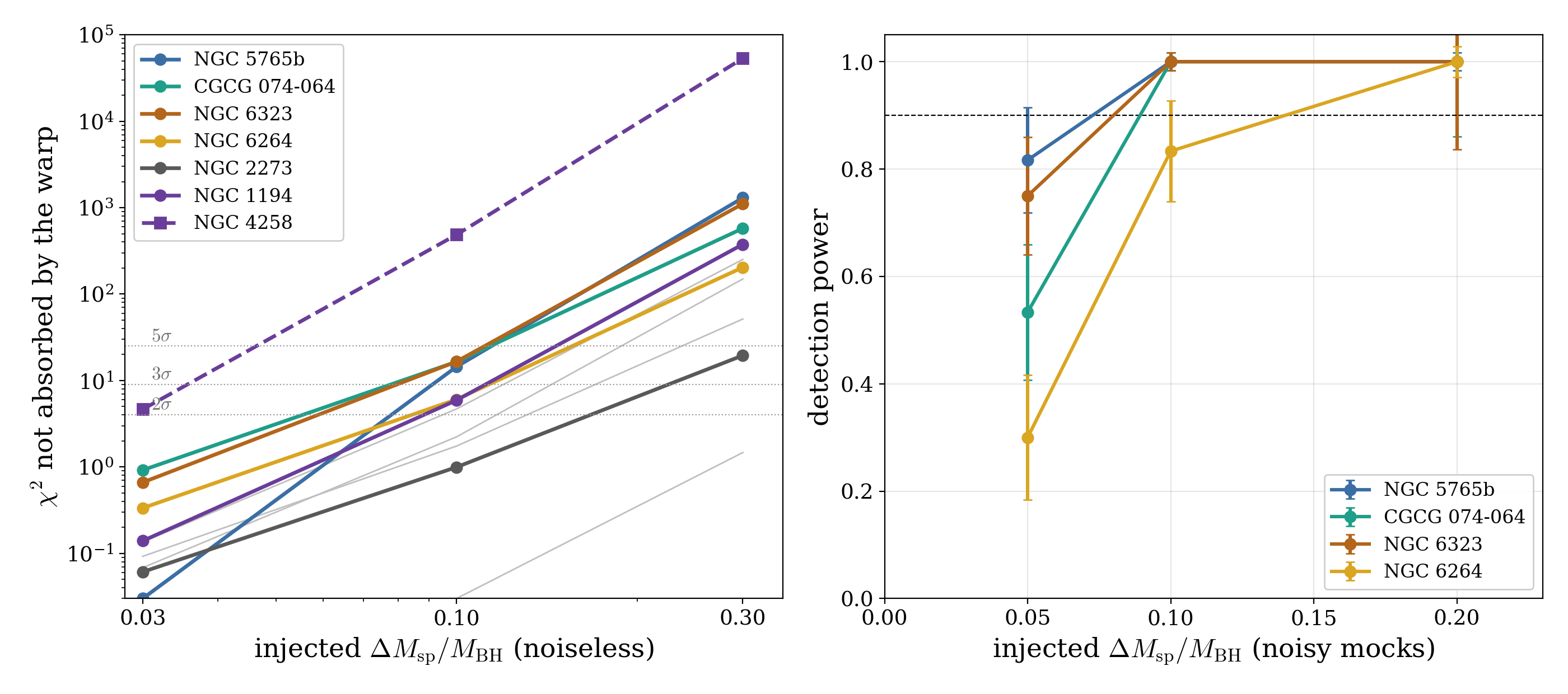}
\caption{Sensitivity of the search. Left: the $\chi^2$ that a noiseless spike of slope $2.25$ leaves behind after a warped point mass has been refitted to it, for spikes adding 3, 10 and 30 per cent of the black-hole mass across each disk's annulus; coloured lines are the four disks in which a ten-per-cent spike is separable from a warp, NGC~2273 (dark grey), and the two disks with $\Delta{\rm BIC}>0$ in the unwarped fit, NGC~1194 and NGC~4258 (purple; dashed for NGC~4258), and light grey lines the other disks. Dotted lines mark the $2$, $3$ and $5\sigma$ equivalents. Right: the fraction of noisy mock disks in which an injected spike is detected above the 95th percentile of the spike-free distribution, for the four identifiable disks, with 100 spike-free and 60 injected realisations per point; the dashed line marks 90 per cent. The real data give no improvement in NGC~5765b, CGCG~074-064 and NGC~6323, and $\dlnl=6.5$ in NGC~6264.\label{fig:sensitivity}}
\end{figure*}

\end{document}